\documentclass[10pt,a4paper]{article} 
\usepackage{amssymb}
\usepackage{graphicx}

\usepackage{url}
\urldef{\mailsa}\path|{jnavarrob@asu.edu|

\begin{document}

\newcommand{\mean}[1]{\left\langle #1 \right\rangle}

%\mainmatter  % start of an individual contribution

\title{Formation of Common Investment Networks by Project Establishment between Agents}

%\titlerunning{Formation of Common Investment Networks}
%
\author{Jes\'us Emeterio Navarro-Barrientos\\
School of Mathematical and Statistical Sciences\\
Arizona State University, Tempe, AZ 85287-1804, USA}

\date{November, 2010}

%\authorrunning{J.-E. Navarro-Barrientos}
%
%\email{jnavarrob@asu.edu}

%\toctitle{Formation of Common Investment Networks by Project Establishment between Agents}

%\tocauthor{Jes\'us -Emeterio Navarro-Barrientos}

\maketitle              % typeset the title of the contribution

\begin{abstract}
We present an investment model integrated with trust-reputation mechanisms where agents interact with each other to establish investment projects. 
We investigate the establishment of investment projects, the influence of the interaction between agents in the evolution of the distribution of wealth, as well as the formation of common investment networks and some of their properties.
Simulation results show that the wealth distribution presents a power law in its tail.
Also, it is shown that the trust and reputation mechanism presented leads to the establishment of networks among agents, which present some of the typical characteristics of real-life networks like a high clustering coefficient and short average path length.
\footnote[1]{J.-Emeterio Navarro-Barrientos, “Formation of Common Investment Networks by Project Establishment between Agents”, in J. Salerno, S. Yang, D. Nau and S.-K. Chai (Eds.): Social Computing and Behavioral-Cultural Modeling and Prediction - SBP 2011, LNCS 6589, Springer (2011) pp.172-179.}\\
\noindent

\end{abstract}
\section{Introduction}

Recently, different socio-economical problems have been modeled using agent-based simulations, presenting a different perspective (usually more flexible and realistic) for modeling social and economical behavior. 
Many important contributions to this field are provided by the research group called \emph{agent-based computational economics (ACE)}~\cite{Tesfatsion02,Tesfatsion06}. 
Different ACE models have been proposed to study for example the relationship between market structure and worker-employer interaction networks \cite{Kirman-Vriend}, investors and brokers in financial markets \cite{LeBaron00}, among others. 
A key concern in many of these studies is to understand the loyalty from buyers to sellers by means of repeated business~\cite{Kirman-Vriend}, as well as the mechanisms for coalition formation between agents which for example may depend on voluntary agreement and payoff of the agents~\cite{Fiaschi-Pacini03}.

Other interesting contributions have been made to understand the emergence of networks between the agents, for example trading networks among buyers and sellers who adaptively select their trade partners by looking at their past experiences~\cite{Tesfatsion02}. 
Finally, also of interest is to investigate the topology of the networks emerging from the interaction between agents, usually using methods borrowed from the field of statistical physics \cite{Albert-Barabasi01}.

The main goal of this paper is to improve the understanding of two main problems in ACE models: (i) the economical component that describes the dynamics of the wealth distribution among agents; and (ii) the social component that describes the dynamics of loyalty, trust and reputation among agents.
For this, we integrate in this article a wealth distribution model based on constant proportional investments and a network formation model where agents interact with each other to establish investment projects. 
%Thus, the following two research questions are investigated in this contribution: 
%\emph{how does the interaction between agents influence the evolution of the distribution of wealth and the trust-reputation between the agents? How does the topology of the networks is characterized and how does it evolve over time?} 
%For this, Section~\ref{sec:wealthNetworks} presents the model for the formation of common investment networks; Section~\ref{sec:portfolioresults} presents some simulation results showing the evolution of the distribution of wealth as well as the evolution of trust and reputation among agents; Section~\ref{sec:commoninvestmentnetworks} investigates the formation of networks for different parameter values in the model and analyzes some of the properties of the networks; and finally in Section~\ref{sec:portfolioconclusions} some conclusions and extensions for these investigations are discussed.

\section{The Model}
\label{sec:wealthNetworks}

\subsection{Wealth dynamics}
Consider an \emph{agent-based system} populated with $N$ agents, where each agent posses a \emph{budget} $x_{k}(t)$ (measure of its ``wealth'' or ``liquidity'') that evolves over time given the following dynamic:
\begin{equation}
  \label{wealthNetworks1}
  x_{k}(t+1)=x_{k}(t)\Big[1 + r_{mk}(t)\,q_{k}(t)\Big] +a(t),
\end{equation}
where $r_{mk}(t)$ denotes the return on investment (RoI) that the agent $k$ receives from its investment $q_{k}(t)$ in project $m$. 
$q_{k}(t)$ denotes a \emph{proportion of investment}, i.e the fraction or ratio of the budget of agent $k$ that the agent prefers to invest in a market and $a(t)$ denotes an external income.

In this model, agent $k$ invests a portion $q_{k}(t)x_{k}(t)$ of its total budget at every time step $t$ yielding a gain or loss in the market $m$, expressed by $r_{mk}(t)$.
Similar wealth models have been presented in \cite{solomon01PhysicaA,Gusman05,NavarroPhysicaA} where the dynamics of the investment model are investigated using some results from the theory of multiplicative stochastic additive processes \cite{Brandt86,Sornette-Cont97}.

Note that this approach assumes that the market, which acts as an \emph{environment} for the agent, is not influenced by its investments, i.e. the returns are exogenous and the influence of the market on the agent is simply treated as random.
This is a crucial assumption which makes this approach different from other attempts to model real market dynamics, e.g. in financial markets \cite{LeBaron00}.

\subsection{Trust-reputation mechanisms and project establishment}

In order to launch a particular investment project $m$ at time $t$, a certain minimum amount of money $I_{\mathrm{thr}}$ needs to be collected among the agents.  
The existence of the investment threshold $I_{\mathrm{thr}}$ is included to enforce the interaction between agents, as they need to \emph{collaborate} until the following condition is reached: 
\begin{equation} 
  \label{threshold} 
  I_{m}(t)=\sum_{k}^{N_{m}} q_{k}(t)\,x_{k}(t) \geq I_{\mathrm{thr}}, 
\end{equation} 
where $N_{m}$ is the number of agents collaborating in the particular investment project $m$.  
There may be different investment projects $m$ at the same time, but for simplicity, it is assumed that each agent participates in only one investment project at a time. 

The first essential feature to be noticed for the formation of common investment networks is the establishment of preferences between agents.
It is assumed that the decision of an agent to collaborate in a project will mainly depend on the previous history it has gained with other agents.
Consider an agent $k$ which accepts to collaborate in the common investment project $m$ initiated by agent $j$.
Thus, agent $k$ receives the following payoff at time $t$:
\begin{equation} 
  \label{payoff} 
  p_{kj}(t)=x_{k}(t)\,q_{k}(t)\,r_{m}(t).
\end{equation} 
Reiterated interactions between agent $k$ and agent $j$ lead to different payoffs over time that are saved in a decision weight: 
\begin{equation} 
  \label{weight2} 
  w_{kj}(t+1)= p_{kj}(t)+ w_{kj}(t)\,e^{-\gamma},
\end{equation} 
where $\gamma$ represents the memory of the agent with initial condition $w_{kj}(0)=0$.

The payoffs obtained from previous time steps $t$ may have resulted from the collaborative action of different agents, however, these are unknown to agent $k$, i.e agent $k$ only realizes the initiator of the project, agent $j$.  
Furthermore, in order to mirror reality, it is assumed that there are more investors than initiators of projects. 
For this, we consider that from the population of $N$ agents only a small number $J$ are initiators, i.e. $J\ll N$, where the reputation of an initiator $j$ can be calculated as follows (for more on trust and reputation models see \cite{Sabater05}):
\begin{equation} 
  \label{eq:reputation} 
  W_j(t+1)= \sum_{k=0}^{N} w_{kj}(t); \quad   W_k(t+1)= \sum_{j=0}^{J} w_{kj}(t).
\end{equation} 

At every time step $t$ an initiator is chosen randomly from the population and assigned with an investment project.
The initiator randomly tries to convince other agents to invest in the project until an amount larger than the threshold $I_{\mathrm{thr}}$ has been collected. 
For this, we use a Gibbs or Boltzmann distribution to determine the probability that the contacted agent $k$ may accept the offer of agent $j$: 
\begin{equation} 
  \label{eq:accept} 
  \tau_{kj}(t)=\frac{e^{\beta w_{kj}(t)}}{ 
    \sum_{i=1}^{J}e^{\beta w_{ki}(t)}},
\end{equation} 
where in terms of the weight $w_{kj}$, the probability $\tau_{kj}(t)$ considers the good or bad previous experience with agent $j$ with respect to the experience obtained with other initiators; and $\beta$ denotes the greediness of the agent, i.e. how much importance does the agent give to the decision weight $w_{kj}$.
%The term $\beta$ is typically called \emph{temperature} and in this case for small $\beta$ values (for example $\beta=0.001$), initiators are equally probable to be chosen. 
%In terms of the behavior of the agent this means that the agent decides to explore more the payoffs that the initiators can offer by not taking into account the previous obtained payoffs.
%For large $\beta$ values (for example $\beta=1$), previous experience with the initiators is enhanced, i.e. the agent decides to explode more those initiators supplying the largest positive payoffs.
In order to take a decision, agent $k$ uses a technique analogous to a roulette wheel where each slice is proportional in size to the probability value $\tau_{kj}(t)$.
Thus, agent $k$ draws a random number in the interval $(0,1)$ and accepts to invest in the project of agent $j$ if the segment of agent $j$ in the roulette spans the random number.
Finally, an initiator $j$ stops to contact other agents if either the investment project has reached the threshold $I_{\mathrm{thr}}$ or if all agents in the population have been asked for collaboration.
%Finally, note that the initiator will always invest in its own project, i.e. neither looks at its own performance  nor compares it with its experience with other initiators.
If the project could be launched it has to be evaluated.  
The evaluation should in general involve certain ``economic'' criteria that also reflects the nature of the project. 
However, for simplicity we assume that the failure or success of an investment project $I_{m}$ is randomly drawn from a uniform distribution, i.e. $r(t)\sim U(-1,1)$.
%A more realistic assumption would include also gains with $r\gg1$, while the loss is still bound to the maximum investment value. 
 
%%%%%%%%%%%%%%%%%%%%%%%%%%%%%%%%%%%%%%%%%%%%%%%%%%%% 
\section{Results of Computer Simulations} 
\label{sec:portfolioresults} 

We performed some simulations using the parameter values in Table~\ref{table:portfolioPars}, for simplicity, we assume that the initial budget is the same for all agents.
Moreover, the proportion of investment is assumed to be constant and the same for all agents i.e. $q_{k}(t)=q=const.$.
\begin{table}[ht]
  \caption{Parameter values of the computer experiments for the investment networks formation model.}
  \label{table:portfolioPars} 
  \begin{center}
    \begin{tabular}{l|l} 
      \hline 
      Parameter & Value \\
      \hline
      Num. of agents & $N=10^{4}$ \\
      Num. of initiators & $J=100$ \\
      Num. of time steps & $t=10^{5}$ \\
      Investment threshold & $I_{\mathrm{thr}}=9$ \\
      Return on Investment & $r\sim U(-1,1)$ \\
      \hline
    \end{tabular}
    \hspace{1cm}
    \begin{tabular}{l|l} 
      \hline 
      Parameter & Value \\
      \hline
      Initial budget & $x_{k}(0)=1$ \\
      Income & $a_{k}=0.5$ \\
      Memory & $\gamma_{k}=0.1$ \\
      Greediness & $\beta_{k}=1$ \\
      \hline
    \end{tabular}
  \end{center}
\end{table} 

\begin{figure}[ht]
  \centering
    \includegraphics[width=6cm]{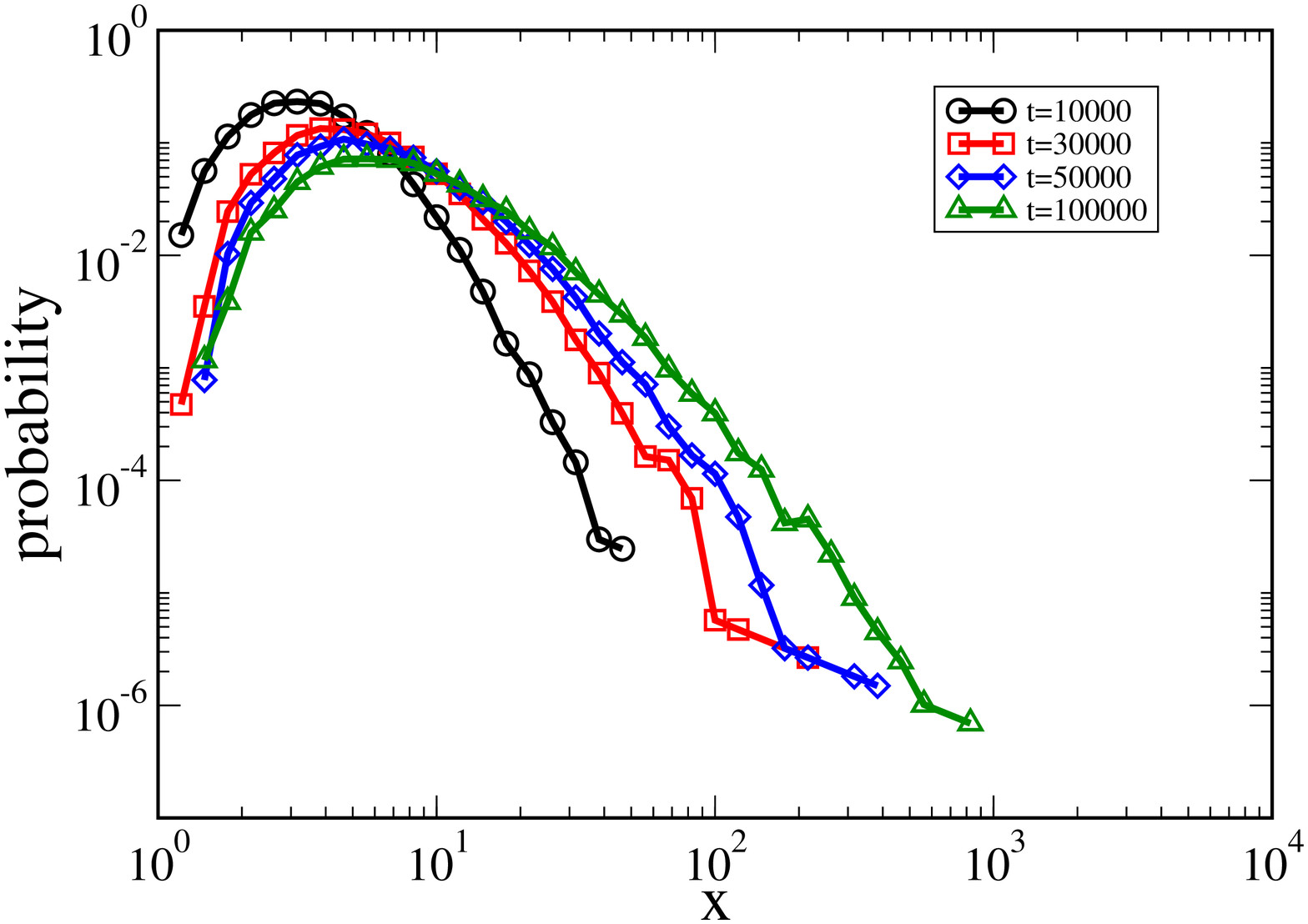}
    \includegraphics[width=6cm]{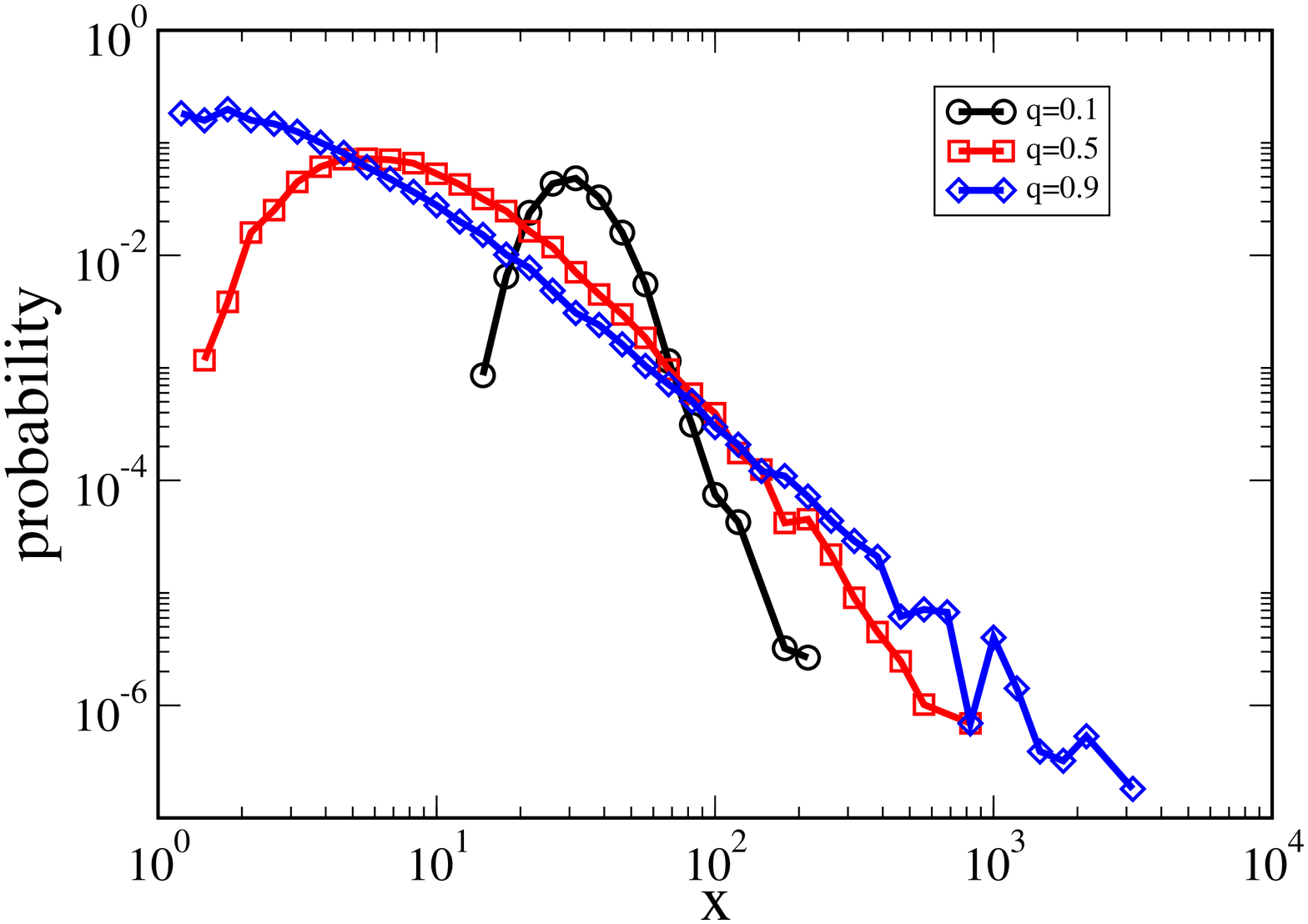}
    \caption{(left) Evolution of the budget distribution over time for $q=0.5$; (right) budget distribution at time step $t=10^{5}$ for different proportion of investment $q$. Additional parameters as in Table~\ref{table:portfolioPars}.}
    \label{fig:net_DistX}
\end{figure}

Fig.~\ref{fig:net_DistX} (left) shows the evolution of the budget distribution over time for $q=0.5$.
Note that the probability distribution of the budget converges to a stationary distribution with a power law in the tail, a property of investment models based on multiplicative processes repelled from zero \cite{Sornette-Cont97,NavarroPhysicaA}.
%Note, however, that a large number of time steps (established projects) are needed to reach a stationary wealth distribution, basically because agents are not always establishing investment projects, i.e. the investments are not larger than the investment threshold $I_{\mathrm{thr}}$ needed to establish a project.

Fig.~\ref{fig:net_DistX} (right) shows the distribution of the budget at time step $t=10^{5}$ for different proportion of investment $q_{k}=q$.
Note that even for a large number of time steps, the budget distribution for agents with a proportion of investment of  $q=0.1$ has not yet converged to a stationary distribution, whereas for $q=0.5$ and $q=0.9$, the distribution reached a stationary state after $t=70000$ and $t=50000$ time steps, respectively.

\begin{figure}[ht]
  \centering
    \includegraphics[width=6cm]{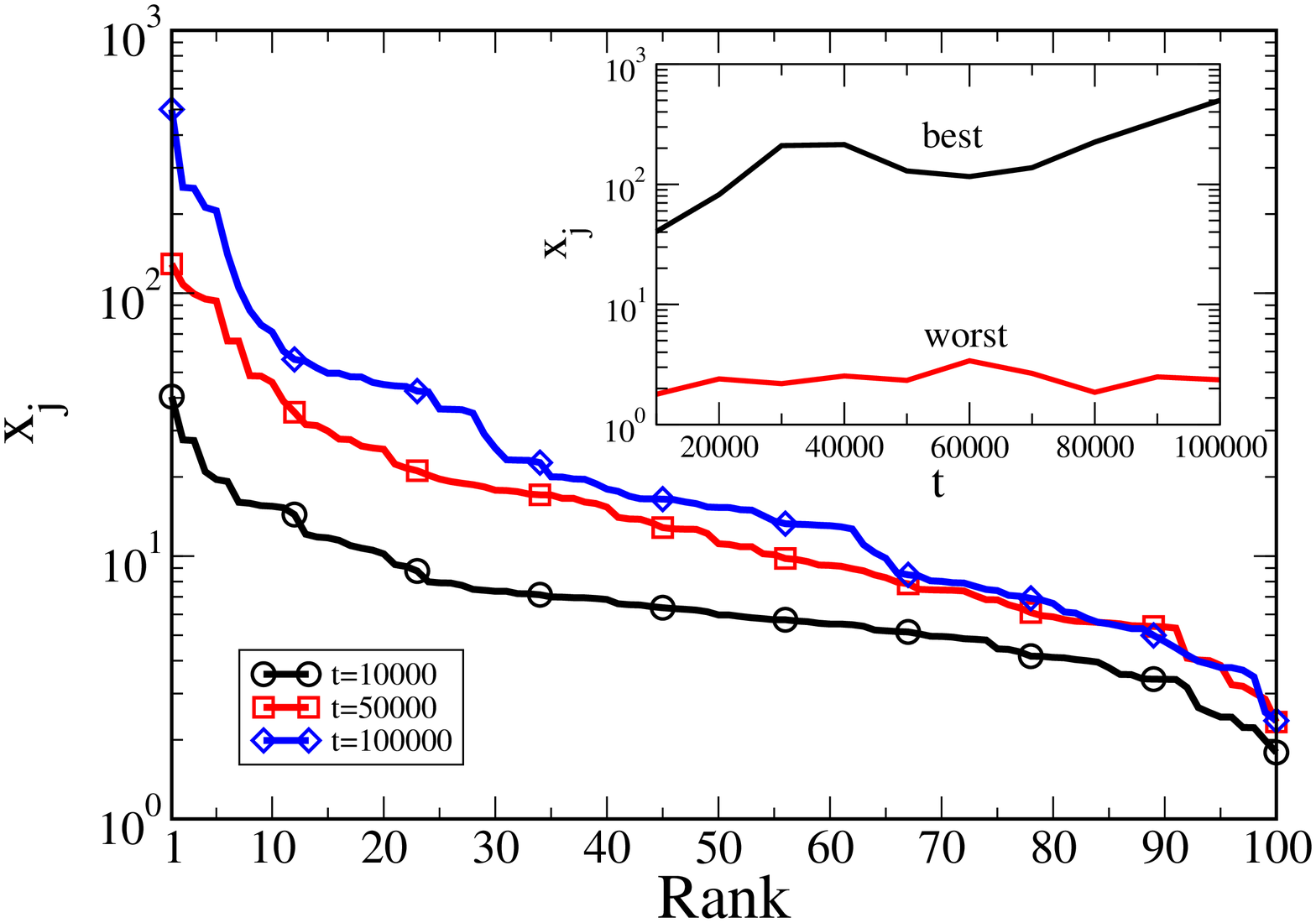}
    \includegraphics[width=6cm]{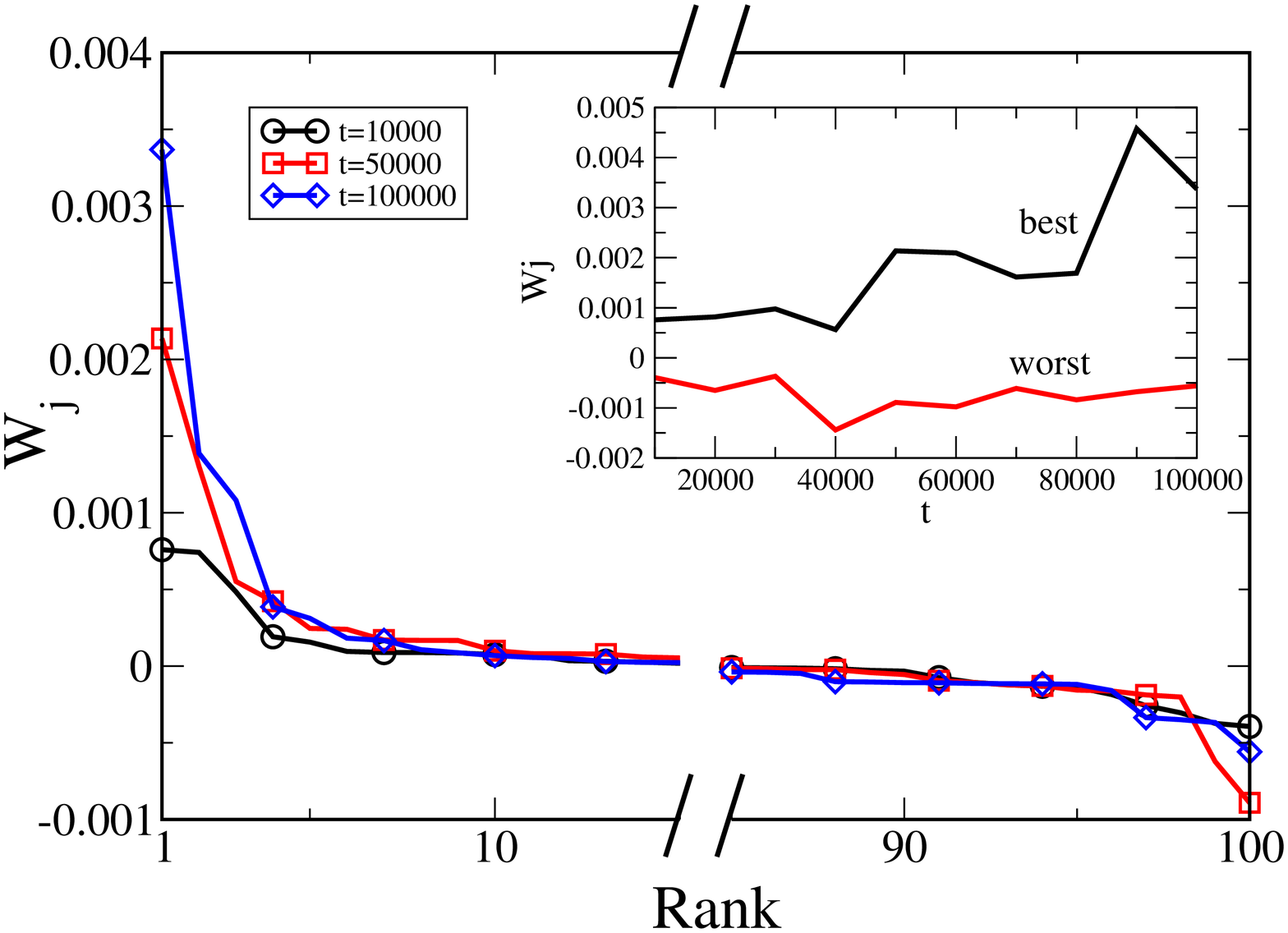}
    \caption{Evolution of the rank-size distribution of initiators for: (left) budget; (right) reputation. Insets show the budget and reputation, respectively, of the best and the worst initiator. Additional parameters as in Table~\ref{table:portfolioPars}.}
    \label{fig:net_BudRep}
\end{figure}

Now, in order to understand the role that initiators play in the dynamics of the investment model, we examine the evolution of their budget and reputation over time.
For the sake of clarity, we show the rank-size distribution of the budget instead of the probability distribution of the budget.
Fig.~\ref{fig:net_BudRep} (left) shows the rank-size distribution of the budget of the initiators, note that the slope of the distribution increases over time.
It is also interesting to examine the evolution of the budget of the initiator with the largest and the smallest budget at the end of the simulation.
This is shown in the inset of Fig.~\ref{fig:net_BudRep} (left), note that the budget of the best agent was not always increasing over time.
%It would be interesting to show the influence of successful and non-successful projects at the beginning of the simulation on the evolution of the budget over time.
%However, because of the fact that initiators randomly ask other agents to invest in their project, they do not really have a preference over wealthy or non-wealthy agents, therefore, it is only the reputation of the initiators the property of the initiator that plays an important role in the dynamics of the investment model.
Fig.~\ref{fig:net_BudRep} (right) shows the rank-size distribution of the reputation of the initiators, Eq.~(\ref{eq:reputation}), note that the distribution does not change over time and only for a small number of agents there is a significant increase or decrease on reputation over time.
Moreover, it can be shown that the average value of the reputation has a shift to larger positive values over the course of time.
This occurs due to aggregation over time of the external incomes $a(t)$ in Eq.~(\ref{wealthNetworks1}) into the dynamics of the decision weights in Eq.~(\ref{weight2}).
Moreover, the inset in Fig.~\ref{fig:net_BudRep} (right) shows the reputation of the best and the worst initiator indicating the presence of no symmetrical positive/negative reputation values.

The influence of the other parameters in the dynamics of the model was also analyzed, however, for the sake of brevity, we discuss only the role of the number of initiators $J$ in the dynamics.
It can be shown that if a less number of initiators is considered, then more investors will be willing to invest in their projects, leading to a larger amount of investment that can be collected by the initiators.
%In order to let a large number of agents to invest in a project from the very beginning of the simulation, the investment threshold $I_{\mathrm{thr}}$ has to be proportional to the initial amount of money that the initiators may be able to collect.
It was mentioned before that the tail of the wealth distribution has a power law distribution and it can be shown that the larger $J$ the larger the slope of the power law. 
The reason for this is that a small number of initiators collect more money from the investors leading to larger profits and looses which over time lead to wider distributions than for a large number of initiators.
%Finally, it can be shown that the initiators tend to accumulate more budget than the rest of the agents
%This occurs because by definition, initiators invest more frequently than those that are no-initiators.

\section{Structure of Common Investment Networks}  
\label{sec:commoninvestmentnetworks}

\begin{figure}[ht]
  \centering
  \includegraphics[width=6cm]{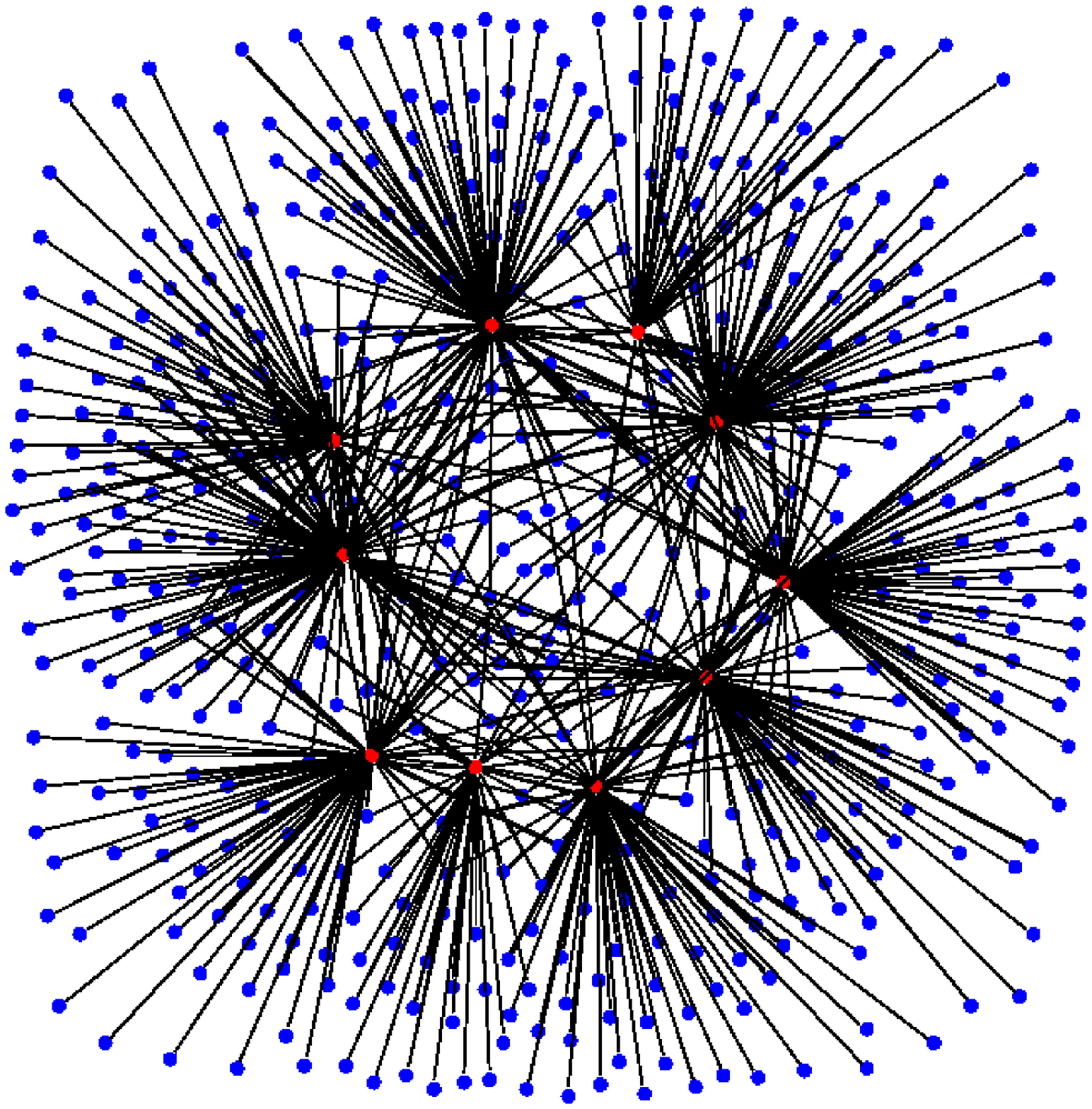}
  \includegraphics[width=6cm]{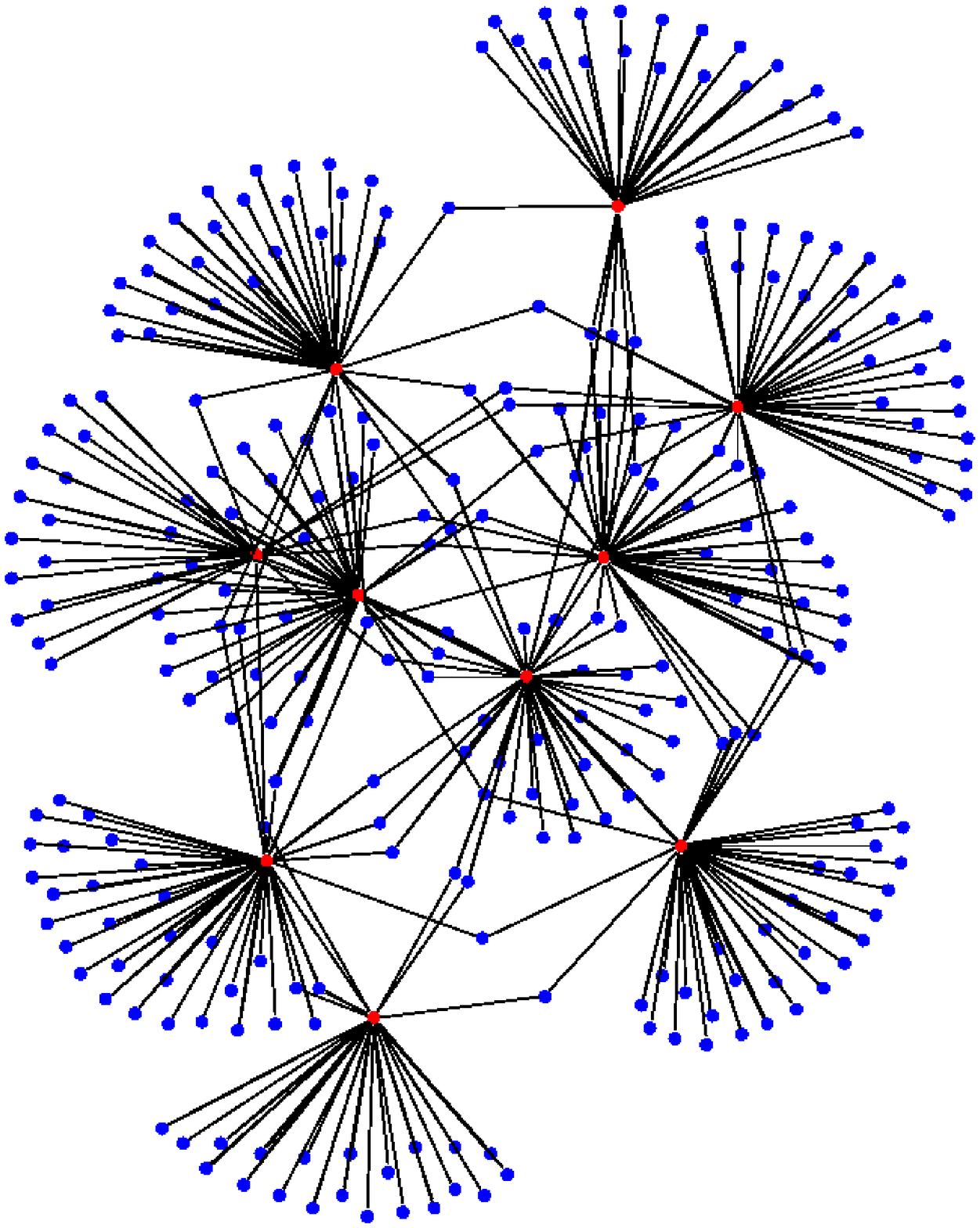}

  \includegraphics[width=7cm]{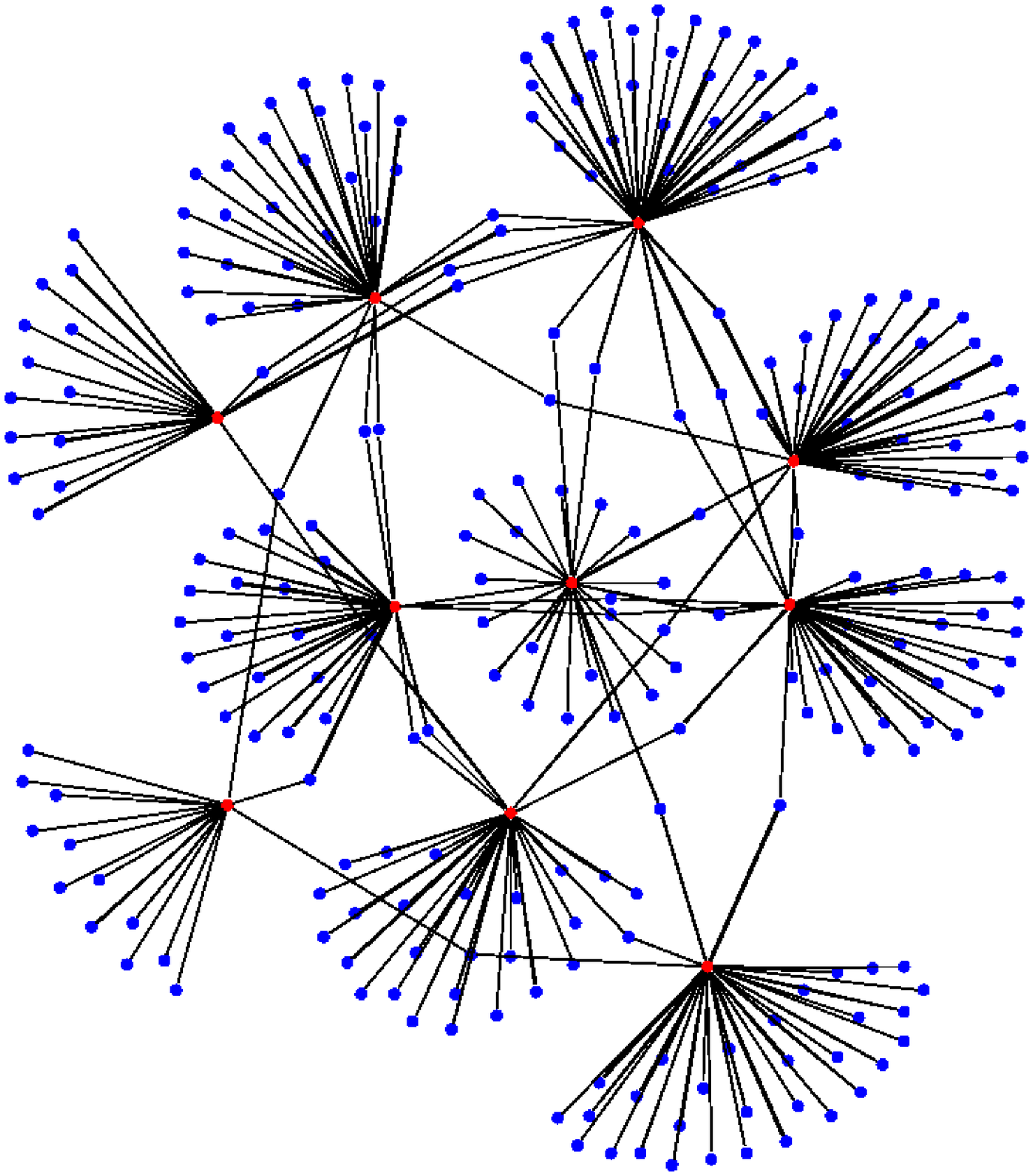}
  \caption{Common investment networks for different proportions of investment at time step $t=1000$: (left) $q=0.1$, (right) $q=0.5$ and (bottom) $q=0.9$. A link between agents represents a positive decision weight, i.e $w_{kj}>0$. For $N=1000$ investors (blue-gray nodes) and $J=10$ initiators (red-bold nodes). Additional parameters as in Table~\ref{table:portfolioPars}.}
  \label{fig:netDiffq}
\end{figure}

In this section, we analyze the topology of the networks for different constant proportion of investment.
For this, we run different computer experiments for a small population of agents $N=1000$ ($N$ is also the number of nodes in the network) and the other parameter values as in Table~\ref{table:portfolioPars}. 
The first experiment investigates the influence of the proportion of investment in the properties of the network.
Fig.~\ref{fig:netDiffq} shows the networks emerging from the investment and trust-reputation models for different proportion of investment $q$ at time step $t=1000$.
Note that these networks have two types of nodes, the red(bold) nodes represent investors and the blue(gray) nodes represent initiators.
Based on visual impression, the density of the network decreases with respect to the proportion of investment.
This occurs because agents investing more also tend to loose more, which leads to more mistrust.

However, from the visual representation of the network it is not possible to draw many conclusions from the dynamics of the networks.
Thus, we obtained the following typical properties of the networks:
\begin{itemize}
\item Number of links $V$ and maximal degree $k_{\mathrm{max}}$ (the degree of the highest degree vertex of the network).
\item Average path length $l$: the average shortest distance between any pair of nodes in the network.
%, i.e. the average of the minimum number of links that are needed to cross from one node to another node. 
Small world networks have a small average path length which scales logarithmically with the size of the network, i.e. $l\sim \log{N}$ \cite{Watts98}.
\item Clustering coefficient $C$: measures the transitivity of the network. It has been shown that in real social networks the clustering coefficient is usually much larger than the clustering coefficient in a random network with the same number of nodes and links \cite{Watts98}.
\end{itemize}

For the sake of brevity, we present on the following the most important results for our analysis on the previous listed properties of the networks.
First, it can be shown that the number of links $V$ over time fits a power law where the slope of the power law decreases if the proportion of investment increases.
%As expected, the number of links increases much faster over time for larger proportion of investment $q$, however, the number of links for small $q$ is much larger in the beginning of the simulations than for larger $q$.
%This occurs because in the beginning the decision weights are zero and their values are much more modified by the payoffs than by the previous weight.
%We note also that the larger $q$, the larger the number of links in the network.
The  maximal degree  $k_{\mathrm{max}}$ of the network also increases over  time with a power-law behavior.
%Based on visual impression, after  $t=10^{5}$ time steps  the number of links  and the maximal degree have a similar value for large proportions of investment.   
%This occurs mainly because in the long run, the decision weights of some agents are largely modified by the previous decision weights than by their current payoff.
It can be shown that the clustering coefficient $C$ is larger for small proportions of investment.
This occurs because a small proportion of investment leads to a higher clustering in the network due to the mistrust that large losses generate in the investors.
%Finally, we noted that for both initiators and investors their degree distributions follows a binomial distribution.
%First studied by \cite{Erdoes-Renyi}, random graphs show the property of binomial degree distribution, which for large number of nodes can be good approximated by a Poisson distribution.

Table~\ref{tab:netNI} shows some of the most important characteristics for different number of investors $N$ and initiators $J$ for a large number of time steps, i.e. $t=10^5$. 
For each network we indicate the average degree $\mean{k}$ (the first moment of the degree distribution), the average path length $l$ and the clustering coefficient $C$.
For comparison reasons we include the average path length $l_{\mathrm{rand}}=\log{(N)}/\log{(\mean{k})}$ and the clustering coefficient $C_{\mathrm{rand}}=\mean{k}/N$ that can be obtained from a random network with the same average degree $\mean{k}$ of the investment networks.
\begin{table}[ht]
  \centering 
  \caption{Properties of the investment networks for different number of investors and initiators. 
    For each network the properties measured are: the average degree $\mean{k}$, the average path length $l$ and the clustering coefficient $C$. For proportion of investment $q=0.5$, $t=10^{5}$ and further parameters as in Table~\ref{table:portfolioPars}.} 
  \label{tab:netNI} 
  \begin{tabular}{c|c|c|c|c|c|c|c|c} 
    \hline 
    $N$ & $J$ & $V$ & $k_{\mathrm{max}}$ & $\mean{k}$ & $l$ & $C$ & $l_{\mathrm{rand}}$ & $C_{\mathrm{rand}}$\\
    \hline
    1000 & 10 & 4847 & 517 & 0.9694 & 2.05766 & 0.74557 & - & 0.0009694\\
    2000 & 20 & 19972& 1050 & 3.9944 & 1.99365 & 0.71337 & 5.488 & 0.0019972\\
    3000 & 30 & 41073 & 1475 & 8.2146 & 1.99314 & 0.71130 & 3.8018 & 0.0027382\\
    10000 & 100 & 134279 & 1477 & 26.86 & 2.1563 & 0.24136 & 2.7989 & 0.002686\\
    \hline 
  \end{tabular}
\end{table}

It can be seen that the average degree $\mean{k}$ increases with respect to the system size. 
%This occurs together with a shift to larger positive values of the degree distribution over time.
Note that for the parameters: $N=1000; J=10$, the average degree of the network is less than one, which means that the network has either trees or clusters containing exactly one link. 
In general, the networks show a small average path length $l\approx 2$, meaning that any investor or initiator in the network is in average connected to each other by two links.
Moreover, for a large number of nodes, the average path of the networks is approximately equal to that from a random graph generated with same average degree of the investment network.
On the other hand, the clustering coefficient of the investment networks is larger than the clustering coefficient of a random network, this indicates the presence of transitivity in our networks.
This occurs mainly because of the large number of investors connected to initiators.
Note that the values of $C$ in our networks are similar to the clustering coefficient obtained for real bipartite networks, for example it has been reported that the clustering coefficient for the network of movie actors is $C=0.79$ \cite{Watts98}.
Note that a property of random networks is that the clustering coefficient decreases with respect to the size of the network.
Finally, note that the clustering coefficient of the networks decreases with respect to $N$, this is in qualitative agreement with properties of small-world networks~\cite{Watts98}.

\section{Conclusions and Further Work} 
\label{sec:portfolioconclusions} 

The most important conclusions that can be drawn from the model here presented are that the budget of the agents reaches a stationary distribution after some time steps and presents a power law distribution on the tail, property discussed in other investment models \cite{Sornette-Cont97,solomon01PhysicaA,NavarroPhysicaA}.
It was shown that the topology of the investment networks emerging from the model was analyzed showing that the networks present some of the typical characteristics of real-life networks like a high clustering coefficient and short average path length.
It was also observed that the evolution over time of the number of links $V$, the maximal degree of the network $k_{\mathrm{max}}$ and the clustering coefficient $C$ can be described by a power-law.

We focused our investigations on the feedback describing the establishment and reinforcement of relations among agents and initiators, which dynamic is mainly driven by the decision weights $w_{kj}(t)$, Eq.~(\ref{weight2}).
This is considered a ``social component'' of the agents' interaction and it was shown how this feedback process based on positive or negative experience may lead to the establishment of networks among agents. 

For simplicity, we have just assumed a random selection of failure or success, but we note that more elaborated economic assumptions, such as market dynamics based on supply and demand, can be taken into account as well.
Furthermore, we noted that the external income sources play an important role on the dynamics of reputation and trust among agents.
The results presented indicate that an extra mechanism or behavioral component needs to be added to the model in order to obtain networks with a stationary power-law degree distribution, property which is usually found in real-world networks.

We note also that further experiments are needed for different memory $\gamma$ and greediness $\beta$ values to understand the influence of these parameters in the dynamics of the networks.

% It would be interesting to analyze the role of the memory $\gamma$ and greediness $\beta$ which describe the exponential decay of the past experience and the importance of the weights $w_{kj}(t)$ in the decision process respectively.
% For example, without memory, i.e. $\gamma\to \infty$, it is expected that agents just randomly gather for a certain project which may describe the \emph{random} scenario.  
% On the other hand, if agents' memory is too long, i.e. $\gamma\to 0$, it is expected that any positive and negative experience will last forever and changes in the structure of the networks may hardly be observed, which may describe the \emph{frozen} scenario. 

% A behavioral component could be also included in order to account for the failure or success of the previous investments and to modify the investment proportion $q(t)$ of the agent accordingly. 

%\bibliographystyle{splncs03}
%\bibliography{bibliography}

%%%%->
%
% ---- Bibliography ----
%
%% \begin{thebibliography}{5}
%% %
%% \bibitem {clar:eke}
%% Clarke, F., Ekeland, I.:
%% Nonlinear oscillations and
%% boundary-value problems for Hamiltonian systems.
%% Arch. Rat. Mech. Anal. 78, 315--333 (1982)

%% \bibitem {clar:eke:2}
%% Clarke, F., Ekeland, I.:
%% Solutions p\'{e}riodiques, du
%% p\'{e}riode donn\'{e}e, des \'{e}quations hamiltoniennes.
%% Note CRAS Paris 287, 1013--1015 (1978)

%% \end{thebibliography}
\end{document}